\documentclass[12pt]{iopart}

\usepackage{iopams}

\usepackage{graphicx}

\begin{document}
\title[Optical beam-induced scattering mode of mid-IR laser microscopy\dots]{Optical beam-induced scattering mode of mid-IR laser microscopy:
a method for defect investigation in near-surface and near-interface
regions of bulk semiconductors}

\author{O~V~Astafiev\footnote{E-mail:  ASTF@KAPELLA.GPI.RU.},
V~P~Kalinushkin\ and V~A~Yuryev\footnote{E-mail: VYURYEV@LDPM.GPI.RU.}}

\address{General Physics Institute of the Russian Academy of Sciences,
38, Vavilov Street, Moscow, GSP--1, 117942, Russia}

\begin{abstract}
This paper presents a new technique of optical beam-induced scattering
of mid-IR-laser radiation, which is a special mode of the recently
developed scanning mid-IR-laser microscopy. The technique in its present form
is designed for investigation of large-scale recombination-active
defects in near-surface and near-interface regions of semiconductor wafers.
However, it can be easily modified for the defect investigations in the
crystal bulk. Being in many respects analogous to EBIC, the present technique
has some indisputable advantages, which enable its application for both
non-destructive laboratory investigations and quality monitoring
in the industry.
\end{abstract}

\section{Introduction}
A method of low-angle mid-IR-light scattering (LALS) in combination with
current carrier photoexcitation in the studied sample has been used for the
investigation of large-scale centers of recombination and gluing in the bulk
of semiconducting materials for many years \cite{1}. This method has not been
applied to the investigation of these defects in near-surface layers of semiconductors
until recently, however, although it is evidently promising for both
the investigation of near-surface and near-interface regions of semiconducting
materials and the inspection of ``working'' layers of technological
semiconductor wafers as well as the studies of large-scale recombination-active
defects (LSRDs) directly in ``working'' layers of semiconductor substrates:
in contrast to e.g. SEM in the EBIC mode, it requires practically no special
preparation of substrate surfaces and consequently affects the physical
properties of neither the studied near-surface layers nor the investigated
defects\,\footnote{The only requirement imposed by this method upon a
substrate\,---\,polishing on both sides, while a considerable part of
technological substrates is polished on one side\,---\,seem to be removed
by use of the methods of the laser heterodyne microscopy \cite{2}.};
the method allows one to investigate the interfaces of semiconductors as well as
the surfaces covered with dielectric coatings (without coating removing, until
the substrate is metallized) that, as far as we know, impossible to do
using any of presently existing methods of material investigation and diagnostics.
Nevertheless the first works demonstrating the applicability of LALS with
surface photoexcitation to investigation of LSRDs in near-surface regions
of semiconductors using Ge single crystals as an example were made by us very
recently \cite{3}. A method for visualization of LSRDs presented in the
current paper\,---\,the optical beam-induced scanning low-angle mid-IR-light
scattering technique (OLALS)\,---\,has become a direct logical development
of these works \cite{4}. A scanning dark-field mid-IR-laser microscope
(scanning LALS or SLALS) recently proposed for the investigation of the
large-scale electrically-active defect accumulations (LSDAs) in semiconductors
[5--8] was applied in this technique as a basic instrument.

\section{Experimental details}
\subsection{Optical diagram}
\begin{figure}[t]
\begin{center}
 \includegraphics[scale=3]{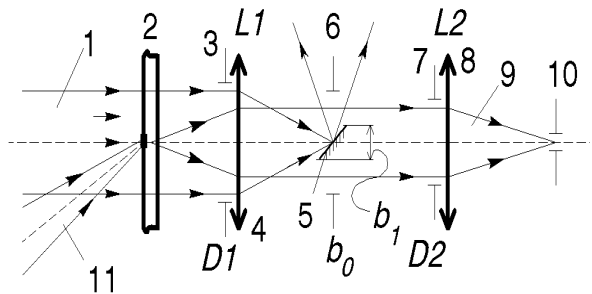}
 \end{center}
\caption{Optical diagram of the SLALS microscope: (1) probe mid-IR-laser beam,
(2) sample, (3,6,7) diaphragms, (4,8) lenses, (9) scattered wave,
(10) IR photodetector, (11) exciting laser beam (used in OLALS
mode).}\label{f1}
\end{figure}

An ideal optical diagram of the SLALS microscope in the OLALS mode (Fig.\,\ref{f1})
gives a sufficiently clear idea of the instrument used in the present work.
Note only the 55-mW He-Ne laser oscillating at the wavelength of 633\,nm
was applied as a light source for the ``surface'' photoexcitation of
the studied samples, the modulated emission of which was focussed on the sample
surface in the back focus of the lens {\sl L1} in the spot with the
dimensions of around 50\,$\mu$m. The imperfection of the optical and
mechanical parts of the laboratory prototype we used impeded us to reach
the optimal focusing of the exciting beam\,\footnote{The microscope resolution
in the OLALS mode is controlled by the effective dimensions of
the domain, in which the non-equilibrium carrier exist, i.e. at the optimal
focusing of the exciting beam the sizes of the light spot must be as small as
possible and not exceed several $\mu$m. The effective dimensions of the
non-equilibrium carrier domain\,---\,and the resolution\,---\,must be controlled only
by the carrier lifetime, diffusion coefficient and the surface recombination
velocity (we mean materials with $\mu$s lifetimes). In addition, the SLALS
microscope itself imposes some restrictions on the sizes of the scattering
domain\,---the domain of non-equilibrium carriers. Due to spacial filtering
applied it depresses the images of too large objects \cite{7,8}.}.
Nonetheless even the available rather imperfect instrument allowed us to
obtain the images of LSRDs in near-surface layers of silicon wafers and demonstrate
the serviceability of the proposed technique.

10.6-$\mu$m CO$_2$-laser emission was used as a probe beam to produce a scattered
wave. A liquid nitrogen cooled MCT photoresistor was used as an IR detector.

To obtain images of LSRDs we used the lock-in detection at the modulation frequency
of the Ne-Ne laser emission (the probe CO$_2$-laser radiation was not
modulated in the OLALS mode), while the lock-in detection at the frequency of
the CO$_2$-laser light modulation was applied to obtain images of the same regions
of the wafers without photoexcitation (in the basic mode of the SLALS microscope).

\subsection{Samples}
76-mm 381$\pm$15\,$\mu$m thick wafers of single crystalline FZ n-Si:P grown in
the $<$100$>$ direction were used as experimental samples. Their specific
resistivity ranged from 16 to 24~$\mu$m. The samples were chemical-mechanically
polished on one side by the manufacturer (``Wacker''), the other side of
the samples was mechanically polished up to the optical precision grade
at GPI of RAS. In addition, we studied a sample of 76-mm CZ n-Si:P grown
in the $<$100$>$ direction (type KEF--4.5) on one side of which 1200\,\AA\  thick
SiO$_2$ layer was produced by oxidation process (this sample was taken from the CCD
matrix production cycle). The opposite side of this sample was polished mechanically
up to the optical precision grade.

\section{OLALS images}
\subsection{LSRDs in subsurface region}
\begin{figure}[t]
\begin{center}
 \includegraphics[scale=1.4]{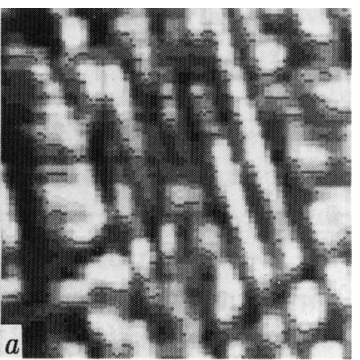}~~
\includegraphics[scale=1.4]{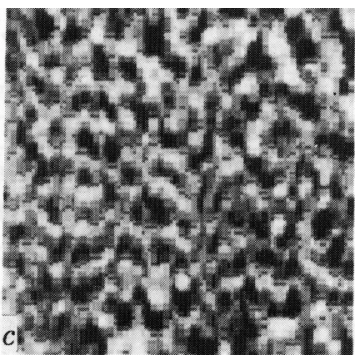}~~
\includegraphics[scale=1.4]{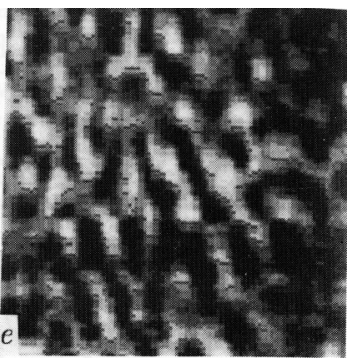}\\
\includegraphics[scale=1.4]{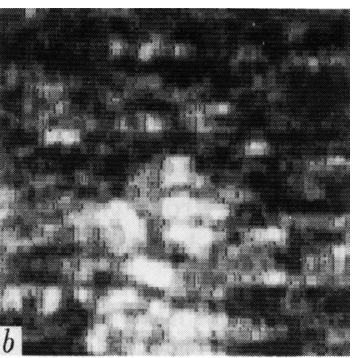}~~
\includegraphics[scale=1.4]{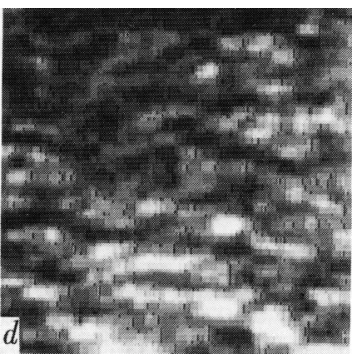}~~
\includegraphics[scale=1.4]{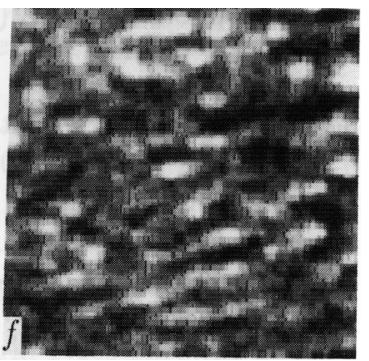}
 \end{center}
\caption{OLALS ({\em a,c,e}\,) and SLALS ({\it b,d,f}\,) images of defects in
FZ Si:P ({\it a--d}\,) and CZ Si:P ({\it e,f}\,):
chemical-mechanical polish ({\it a,b}\,),
mechanical polish ({\it c,d}\,), under 1200\,\AA\
thick SiO$_2$ layer ({\em e,f}\,); 1$\times$1\,mm$^2$.}\label{f2}
\end{figure}

Fig.\,\ref{f2}\,$(a)$ presents an OLALS image of the subsurface layer of
chemical-mechanically polished side of one of the wafers.
The same region scanned without photoexcitation (SLALS image) is given in
Fig.\,\ref{f2}\,$(b)$. The average signal level in Fig.\,\ref{f2}\,$(a)$
is 5--7 times greater than that in Fig.\,\ref{f2}\,$(b)$. The image contrast
in Fig.\,\ref{f2}\,$(b)$\,---\,white spots\,---\,is determined by LSDAs in
the crystal bulk (these defects were studied by LALS e.g. in \cite{1,9}),
while the contrast in Fig.\,\ref{f2}\,$(a)$\,---\,dark objects\,---\,is
caused by domains with the enhanced recombination rate in the near-surface
layer of the wafer, i.e. LSRDs. It is seen that most of these LSRD-rich
domains look like traces of scratches but no scratches were observed on this side of
the wafer by conventional methods.

Fig.\,\ref{f2}\,$(c)$ shows an OLALS image of the opposite\,---\,mechanically
polished side\,---\,of the same wafer, and Fig.\,\ref{f2}\,$(d)$ presents
an image of the same region as Fig.\,\ref{f2}\,$(c)$ in the SLALS mode.
The average signal levels in Figs.\,\ref{f2}\,$(b)$ and \ref{f2}\,$(d)$
practically equal, while that in Fig.\,\ref{f2}\,$(c)$ is 10--15 times
lower than in Figs.\,\ref{f2}\,$(b)$ and \ref{f2}\,$(d)$. A chaotic conglomeration
of small LSDAs is seen in Fig.\,\ref{f2}\,$(c)$ in the damaged by mechanical
polishing subsurface layer.

As it should be expected, absolutely different LSRD-rich domains were
observed on the wafer sides subjected to different polish processes.

Note that similar inferences were previously made  by us from the experiments
on LALS with surface photoexcitation of Ge samples \cite{3}. The difference
of the data presented in Ref.\,\cite{3} from the results of this work consists
in the following: the big stripe-like LSRDs observed in the present work on
the sample side subjected to chemical-mechanical polish cannot be revealed from the
light-scattering diagrams in LALS. That is why only defects similar to impurity
clouds \cite{1,9} were observed in Ref.\,\cite{3} (as it was established in
Ref.\,\cite{3}, the latter are also LSRDs). On the mechanically polished
side of the samples, like in the present work, rather intense light scatter
by small LSRDs was registered in the near-surface layer \cite{3}.

Remark also that different schemes of elecron--hole pairs were used
in the present work and in the work \cite{3}. In the current work,
as mentioned above, the non-equilibrium carriers in the near-surface layer
were excited by the focussed light beam in a quasi-continuous regime, and
a whole domain of the non-equilibrium carriers scattered light, while in \cite{3},
the photoexcitation was produced with a wide light beam in a pulse regime and
a surface inhomogeneity in the distribution of non-equlibrium carrier
concentration inside the sight spot caused by subsurface LSRDs was a source
of the mid-IR-light scatter. In both cases, the light scatter intensity
must be proportional to the square of the non-equilibrium carrier
concentration and, in case of linear recombination, the square of
the photoexcitation power. The real excitation power dependances
will be discussed below.

\subsection{LSRDs under SiO$_2$ layer}
Fig.\,\ref{f2}\,$(e)$ presents an OLALS image of the silicon wafer surface
under the oxide layer. The dark spots in the image represent the domains with low
non-equilibrium carrier lifetime\,\footnote{The closely spaced parallel stripes
are the noises brought in the picture by the scanner mechanics.}. As far as we
know, such defects of a near-interface semiconductor layer directly under the
oxide layer have never been observed thus far. It is clear that such a strong
nonuniformity of recombination properties of the subinterface layer must
disastrously affect the quality of devices, in particular CCD matrices, from the
production cycle of which the studied sample was taken. It is obvious also that
these defects arose due to wafer processing before the oxidation.

Fig.\,\ref{f2}\,$(f)$ shows a SLALS image of the same region on the sample
(the white spots are the images of LSDAs in the wafer bulk).
The pattern in this picture is absolutely different from that in
Fig.\,\ref{f2}\,$(e)$ because different imperfections were revealed
in each case.

\subsection{OLALS signal dependence on photoexcitation power}
\begin{figure}[t]
 \begin{center}
 \includegraphics[scale=3]{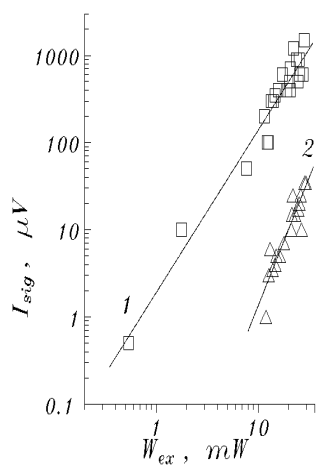}
 \end{center}
\caption{Dependances of IR-photodetector signal on the absorbed power of
the exciting radiation: chemical-mechanical (1) and mechanical (2) polish,
FZ Si:P wafer depicted in Fig.\,2.}\label{f3}
\end{figure}

Fig.\,\ref{f3} demonstrates the dependencies of the scattered
CO$_2$-laser light intensity registered by the detector on the power of
He-Ne laser radiation absorbed by the crystal for the
FZ Si:P wafer sides subjected
to chemical-mechanical (1) and mechanical polish processes, OLALS images
of which are shown in Fig.\,\ref{f2}. The square dependence obtained for
chemical-mechanically polished side verifies the fact that this surface is
imaged in the rays scattered by the non-equilibrium carrier domain and the
recombination is linear\,\footnote{It is generally known that the light
scattering intensity is proportional to the square of the dielectric constant
deviation in the scatterer.}. The situation is much more sophisticated for the
mechanically polished side, for which the cubic  dependence was obtained.
The similar dependence was previously obtained by us for the mechanically
lapped surface of the Ge sample in Ref.\,\cite{3}. As of now, we have got
no perspicuous explanation for these dependances. Perhaps the cubic
dependence is conditioned by the recombination processes in the vicinity
of a huge amount of potential barriers in the damaged by mechanical polish
subsurface layer of the samples \cite{10}. It is obvious, though, that in
this case the surface image is also formed in the rays scattered by the
non-equilibrium carrier domain.

\section{Conclusion}
Summarizing the above we can conclude that this work presents an optical
non-destructive technique which can be a powerful tool for the investigation
of near-surface and near-interface layers of semiconducting materials.
The described method is in some respects analogous in its physics to
EBIC or OBIC but as distinct to the latter ones, it requires neither
special preparation of a sample surface nor Schottky barrier
or {\it p--n} junction.
It enables the investigation of semiconductor interfaces and surfaces
coated with dielectric layers and does not require to remove the coatings.

To demonstrate the method potentialities, the images of LSRDs situated in
the vicinity of the chemical-mechanically and mechanically polished
surfaces of single crystalline silicon were obtained without preliminary
preparation of the surfaces. The images of LSRDs located near the surface,
on which the SiO$_2$ layer was created by the industrial process of
oxidation, were also obtained without removing the coating  and surface
processing. As far as we know, such images can be obtained by means of no
presently used methods of microscopy. The defects observed on the
Si crystal side subjected to chemical-mechanical polish and under the
oxide layer, caused by the technological treatments of the wafers,
undoubtedly must  disastrously affect the serviceability of devices
fabricated of such wafers.

We should remark in the conclusion that the technique described is
easily adaptable for work directly in the technological line of
semiconductor devices production and can serve as a promising tool
e.g. for monitoring of the ``working'' layer quality. Some potential
industrial applications of the technique are disused in Refs.\,\cite{6,8}.
\\

\end{document}